\documentclass{article}
\usepackage{amssymb}
\usepackage{amsmath}
\usepackage{subcaption}
\usepackage{float}
\usepackage{graphicx} 

\newtheorem{Definition}{Definition}[section]

\title{Fractal Schr\"{o}dinger Equation: Implications for Fractal Sets}
\author{Alireza Khalili Golmankhaneh $^1$\footnote{Corresponding Author }, Stergios Pellis $^2$, Massimiliano Zingales $^3$\\
$^1$ Department of Physics, Urmia Branch, \\Islamic Azad University, Urmia, 63896, Iran\\
alirezakhalili2002@yahoo.co.in\\
$^2$ Department of Physics, University of Ioannina, Ioannina, Greece\\
sterpellis@gmail.com\\
$^3$ Dipartimento di Ingegneria, Universit\`{a} degli Studi di Palermo,\\ Viale delle Scienze ed.8, Palermo, 90128, Italy\\
massimiliano.zingales@unipa.it}

\begin{document}

\maketitle

\begin{abstract}
This paper delves into the world of fractal calculus, investigating its implications for fractal sets. It introduces the Fractal Schr\"{o}dinger Equation and provides insights into its consequences. The study presents a General Solution for the Time-Dependent Schr\"{o}dinger Equation, unveiling its core aspects. Exploring quantum mechanics in the context of fractals, the paper analyzes the Probability Density of the Radial Hydrogen Atom, unveiling its behavior within fractal dimensions. The investigation extends to deciphering the intricate Energy Levels of the Hydrogen Atom, uncovering the interplay of quantum mechanics and fractal geometry. Innovatively, the research applies the Fractal Schr\"{o}dinger Equation to Simple Harmonic Motion, leading to the introduction of the Fractal Probability Density Function for the Harmonic Oscillator. The paper employs a series of illustrative figures that enhance the comprehension of the findings. By intertwining quantum mechanics and fractal mathematics, this research paves the way for deeper insights into their relationship.
\end{abstract}





\section{Introduction}
The concept of fractal geometry was pioneered by Benoit Mandelbrot \cite{Mandelbro}. This revolutionary approach explores intricate geometric forms that defy conventional boundaries, exhibiting fractal dimensions surpassing their topological counterparts. These structures, known as fractals, possess inherent self-similarity and frequently exhibit dimensions that transcend integer values \cite{falconer1999techniques,b-6}. Given that fractals possess distinct measures like the Hausdorff measure, conventional metrics such as length, surface area, and volume-typically applied to Euclidean geometric forms-prove inadequate for fractal analysis \cite{ma-12,ord1983fractal,samayoa2022map,Edgar}. \\
Numerous researchers have approached fractal analysis through various methods, including harmonic analysis \cite{ma-8,ma-13}, measure theory \cite{freiberg2002harmonic,Withers,jiang1998some,giona1995fractal,bongiorno2018derivatives,bongiorno2015fundamental,bongiorno2011henstock}, probabilistic approaches \cite{ma-7}, fractional space \cite{stillinger1977axiomatic}, fractional calculus \cite{ma-6}, and unconventional techniques \cite{Nottale-4}.\\
Fractal calculus arises through an extension of conventional calculus, encompassing differential equations whose solutions manifest as functions characterized by fractal support, including fractal sets and curves. This type of calculus is algorithmic in nature and offers a more streamlined approach when compared to alternative methodologies  \cite{parvate2009calculus,parvate2011calculus,Alireza-book}.
Fractal calculus found generalization and application within the realm of physics, extending to stochastic equations and the establishment of Fourier and Laplace transforms \cite{golmankhaneh2021equilibrium,balankin2023vector,golmankhaneh2018sub,golmankhaneh2016diffraction,khalili2019random,banchuin20224noise,banchuin2022noise,golmankhaneh2021fractalBro}.
The resolution of fractal delayed, neutral, and renewal delay differential equations with constant coefficients was accomplished using the method of steps and Laplace transforms \cite{golmankhaneh2023initial}.
Fractal counterparts of Newtonian, Lagrangian, Hamiltonian, and Appellian mechanics were proposed. Fractal $\alpha$-velocity and $\alpha$-acceleration were defined, enabling the formulation of the Langevin equation on fractal curves \cite{GAA}. The fractal Frechet derivative and the fractal generalized Euler-Lagrange equation and the fractal Du Bois-Reymond optimality condition were introduced \cite{KhaliliVariationa}.\\
The Hermiticity of the fractional Hamilton operator was demonstrated, and the conservation law for parity in fractional quantum mechanics was established. Physical applications yield energy spectra for a hydrogen-like atom \cite{laskin2002fractional,ionescu2014nonlinear}.
The space-time fractional Schr\"{o}dinger equation was investigated through the incorporation of the Caputo fractional derivative and quantum Riesz fractional operator \cite{dong2008space}.
The generalized fractional Schr\"{o}dinger equation, incorporating space and time fractional derivatives, was formulated and subsequently solved for both a free particle and a square potential well. The solution techniques involved integral transforms, Fourier transform, and Laplace transform, resulting in expressions of solutions using the Mittag-Leffler function \cite{wang2007generalized}.\\
The spectral attributes of fractal structures were investigated, involving the introduction of the concept of spectral dimensionality through a straightforward scaling approach. Computations of the Schr\"{o}dinger equation spectrum were conducted for a regular fractal structure, specifically the Sierpinski gasket \cite{wang2007generalized}.
The fractional Schr\"{o}dinger equation for a particle confined within an infinite potential well also for delta potentials were studied \cite{luchko2013fractional,de2010fractional,bayin2012consistency}.
A solution to the fractional Schr\"{o}dinger equation for the one-dimensional harmonic oscillator was provided \cite{jeng2010nonlocality}.
The spectrum of the Schr\"{o}dinger equation on a regular fractal structure was explicitly calculated. As illustrative examples, certain physical properties of percolating clusters near the percolation threshold were discussed \cite{rammal1984harmonic,wang2021novel,jumarie2001schrodinger}.
An asymptotic eigenvalue counting function was formulated for the Schr\"{o}dinger operator with unbounded potentials on several types of unbounded fractal spaces \cite{chen2015spectral}.
The Schr\"{o}dinger equation was solved on a variety of fractal lattices using a recursive technique and the resulting energy levels are discrete, very closely spaced, and highly degenerate \cite{domany1983solutions}.
A versatile approach was proposed for generating wave functions with predetermined fractal dimensions across a range of quantum scenarios, such as the infinite potential well, harmonic oscillator, linear potential, and free particle \cite{wojcik2000time}.
Fractal-like photonic lattices were experimentally realized using the cw-laser-writing technique, leading to the observation of distinct compact localized states-associated with different flatbands within the same lattice configuration \cite{xie2021fractal}.
A technique for enhancing the finite element method on the Sierpinski gasket was demonstrated, enabling unrestricted space partitioning. This enhanced method was applied to numerically analyze solutions of the Schr\"{o}dinger equation with well-type potentials, along with the wave equation \cite{coletta2004numerical}.
Dirac operators and magnetic Schr\"{o}dinger Hamiltonians on fractals while demonstrating their (essential) self-adjointness was suggested \cite{hinz2013dirac}.
Regular and irregular sampling for Sierpinski Gasket functions is examined, where "bandlimited" refers to finite expansions in the first  Dirichlet eigenfunctions of the Laplacian \cite{oberlin2003sampling}.
The analogue of the quantum-mechanical harmonic oscillator on infinite blowups of the Sierpinski Gasket is studied using the standard Kigami Laplacian \cite{fan2009harmonic}.
Driven by these motivations, this paper undertakes the generalization of the Schr\"{o}dinger equation onto a fractal framework. Within this context, we successfully addressed the Schr\"{o}dinger equation for fractal Hydrogen atoms and fractal simple harmonic oscillators.\\
The manuscript's structure is outlined as follows:\\
In Section \ref{1g}, we provide an overview of fractal calculus applied to fractal sets. Section \ref{2g} introduces the Schr\"{o}dinger equation in the context of fractal space and time. We proceed to solve this equation for both the Hydrogen atom and the simple harmonic oscillator, determining probability density functions and energy levels, which are then compared to their standard counterparts. Our concluding remarks are presented in Section \ref{3g}.

\section{Fractal calculus for fractal sets \label{1g}}

In this context, we provide a concise overview of the fractal calculus applied to sets $\mathbf{F}$ contained within the interval $[c_{1},c_{2}]$ of the real number line, as described by Parvate et al. (2009) in their work on calculus \cite{parvate2009calculus}.
\begin{Definition}
 The flag function of $\mathbf{F}$ is introduced, defined by:
\begin{equation}
  \rho(\mathbf{F},I)=\left\{
              \begin{array}{ll}
                1, & \textmd{if}~~ \mathbf{F}\cap I\neq\emptyset;\\
                0, & \textmd{otherwise},
              \end{array}
            \right.
\end{equation}
where $I=[c_{1},c_{2}]\subset \mathbb{R}$.
\end{Definition}
\begin{Definition}
 The coarse-grained mass of $\mathbf{F}\cap [c_{1},c_{2}]$ is defined as:
\begin{equation}
\xi_{\delta}^{\alpha}(\mathbf{F},c_{1},c_{2})=\inf_{|\mathfrak{P}|\leq \delta}\sum_{i=0}^{n-1}\Gamma(\alpha+1)(z_{i+1}-z_{i})^{\alpha}
\rho(\mathbf{F},[z_{i},z_{i+1}]),
\end{equation}
where
\begin{equation}
|\mathfrak{P}|=\max_{0\leq i\leq n-1}(z_{i+1}-z_{i}),
\end{equation}
and $0< \alpha\leq1$.
\end{Definition}
\begin{Definition}
 The mass function of $\mathbf{F}$ is established as:
\begin{equation}
\xi^{\alpha}(\mathbf{F},c_{1},c_{2})=
\lim_{\delta\rightarrow0}\xi_{\delta}^{\alpha}(\mathbf{F},c_{1},c_{2}).
\end{equation}
\end{Definition}
\begin{Definition}
 The fractal dimension of $\mathbf{F}\cap [c_{1},c_{2}]$ is defined as follows:
\begin{align}
  \dim_{\xi}(\mathbf{F}\cap [c_{1},c_{2}])&=\inf\{\alpha:\xi^{\alpha}(\mathbf{F},c_{1},c_{2})=0\}\nonumber\\&
=\sup\{\alpha:\xi^{\alpha}(\mathbf{F},c_{1},c_{2})=\infty\}.
\end{align}
In this manner, we succinctly encapsulate the fundamental concepts within the realm of fractal calculus concerning sets $\mathbf{F}$ contained within the interval $[c_{1},c_{2}]$. The definitions encompass the flag function, coarse-grained mass, mass function, and fractal dimension, all of which play pivotal roles in the analysis of these fractal structures.
\end{Definition}
\begin{Definition}
 The integral staircase function associated with $\mathbf{F}$ is introduced, defined as follows:
\begin{equation}
 S_{\mathbf{F}}^{\alpha}(z)=\left\{
                     \begin{array}{ll}
                       \xi^{\alpha}(\mathbf{F},c_{0},z), & if~z\geq c_{0} ; \\
                      - \xi^{\alpha}(\mathbf{F},z,c_{0}), & otherwise,
                     \end{array}
                   \right.
\end{equation}
where $c_{0}\in \mathbb{R}$ is a fixed number.
\end{Definition}
\begin{Definition}
Definition 6: For a function $g$ defined on an $\alpha$-perfect fractal set, the $F^{\alpha}$-derivative of $g$ at $x$ is defined as:
\begin{equation}
  D_{\mathbf{F}}^{\alpha}g(z)=\left\{
                       \begin{array}{ll}
                         \underset{ y\rightarrow z}{\mathbf{F}_{-}lim}~\frac{g(y)-g(z)}
{S_{\mathbf{F}}^{\alpha}(y)-S_{\mathbf{F}}^{\alpha}(z)}, & if~ z\in \mathbf{F}; \\
                         0, & otherwise,
                       \end{array}
                     \right.
\end{equation}
assuming the existence of the fractal limit $\mathbf{F}_{-}\text{lim}$ as described in the work by Parvate et al. \cite{parvate2009calculus}.
\end{Definition}
\begin{Definition}
 The $\mathbf{F}^{\alpha}$-integral of a bounded function $g(z)$, where $g$ belongs to the set $B(F)$ (indicating that $g$ is a bounded function of $F$), is defined as follows:
\begin{align}
\int_{a}^{b}g(z)d_{\mathbf{F}}^{\alpha}z&=\sup_{\mathfrak{P}{[c{1},c_{2}]}}
\sum_{i=0}^{n-1}\inf_{z\in \mathbf{F}\cap I}g(z)(S_{\mathbf{F}}^{\alpha}(z_{i+1})-S_{F}^{\alpha}(z_{i}))
\nonumber\\&=\inf_{\mathfrak{P}{[c{1},c_{2}]}}
\sum_{i=0}^{n-1}\sup_{z\in \mathbf{F}\cap I}g(z)(S_{\mathbf{F}}^{\alpha}(z_{i+1})-S_{\mathbf{F}}^{\alpha}(z_{i})),
\end{align}
where $z\in \mathbf{F}$, and the infimum or supremum is taken over all subdivisions $\mathfrak{P}{[c{1},c_{2}]}$.

In this manner, we elucidate the essential concepts of the integral staircase function, the $\mathbf{F}^{\alpha}$-derivative, and the $\mathbf{F}^{\alpha}$-integral, which play integral roles in fractal calculus applied to sets $\mathbf{F}$. These definitions underscore the mathematical foundation required for analyzing and understanding fractal structures within the specified context.
\end{Definition}
\section{Fractal Schr\"{o}dinger Equation \label{2g}}
In this segment, we apply fractal calculus to modify the Schr\"{o}dinger equations of hydrogen atoms, revealing the manifestation of fractal properties within the established models. This approach allows us to explore the incorporation of fractal concepts in both spatial and temporal dimensions.
\subsection{Fractal Schr\"{o}dinger Equation of Hydrogen Atom }
The time-dependent Schr\"{o}dinger equation for the hydrogen atom within a fractal space characterized by a dimension of $2+\alpha$, along with fractal time characterized by a dimension of $\beta$, can be expressed in its entirety as follows:
\begin{equation}
i\hbar D_{t}^{\beta} \Psi(r, \theta, \phi, t) = \left(-\frac{\hbar^2}{2m_e} \nabla^2 - \frac{e^2}{4\pi\epsilon_0 S_{\mathbf{F}}^{\alpha}(r)}\right) \Psi(r, \theta, \phi, t),~~r, t\in \mathbf{F},
\end{equation}
where $0\leq \theta \leq \pi,~~ 0\leq \phi \leq 2 \pi $.
In polar coordinates with a dimensionality of $2+\alpha$, the Laplace operator (Laplacian) is represented using fractal derivatives in the following manner:
\begin{equation}
\nabla^2 = \frac{1}{S_{\mathbf{F}}^{\alpha}(r)^2} D_{r}^{\alpha} \left(S_{\mathbf{F}}^{\alpha}(r)^2 D_{r}^{\alpha}\right) + \frac{1}{S_{\mathbf{F}}^{\alpha}(r)^2 \sin \theta}\frac{\partial}{\partial \theta}\left(\sin \theta \frac{\partial}{\partial \theta}\right) + \frac{1}{S_{\mathbf{F}}^{\alpha}(r)^2 \sin^2 \theta}\frac{\partial^2}{\partial \phi^2}
\end{equation}
where $D_{r}^{\alpha}$ is the fractal radial derivative operator.
In $2+\alpha$ polar coordinates employing fractal derivatives, the Laplacian comprises three distinct segments: the fractal radial segment, the polar angle segment, and the azimuthal angle segment. These segments correspond to variations occurring in the radial, polar angle, and azimuthal angle directions, respectively.
\section*{General Solution for Time-Dependent Schr\"{o}dinger Equation}
A proposed solution for the time-dependent Schr\"{o}dinger equation, which incorporates fractal dimensions for both time and space in the context of the hydrogen atom, is indicated by:
\begin{align}
\Psi(\mathbf{r}, t) &= \sum_n c_n \psi_n(\mathbf{r}) e^{-\frac{i}{\hbar} E_n^{\alpha} S_{\mathbf{F}}^{\beta}(t)}\nonumber\\
&\propto \sum_n c_n \psi_n(\mathbf{r}) e^{-\frac{i}{\hbar} E_n^{\alpha} t^{\beta}}
\end{align}
where $E_n^{\alpha}$  are the corresponding fractal energy eigenvalues. This solution shows the fractal time evolution of the wave function and how it changes over fractal time for a quantum system described by the Schr\"{o}dinger equation.
The radial part fractal discretional equation of the Schr\"{o}dinger equation for the hydrogen atom with the replaced potential energy term is given by:
\begin{align}\label{uffupo}
-\frac{\hbar^2}{2m} &\left( \frac{1}{S_{\mathbf{F}}^{\alpha}(r)^2} D_{r}^{\alpha} \left( S_{\mathbf{F}}^{\alpha}(r)^2 D_{r}^{\alpha}R_{nl}(r) \right)- \frac{l(l+1)}{S_{\mathbf{F}}^{\alpha}(r)^2} R_{nl}(r) \right)\nonumber\\& - \frac{e^2}{4\pi\epsilon_0} \frac{1}{S_{\mathbf{F}}^{\alpha}(r)} R_{nl}(r) = E_n^{\alpha} R_{nl}(r).
\end{align}
Where $S_{F}^{\alpha}(r)$  is the radial staircase function, \(m\) is the electron mass, $l$  is the angular momentum quantum number, $R_{nl}^{\alpha}(r)$ is the radial wave function, $e$ is the elementary charge, $\epsilon_0$ is the vacuum permittivity, and $E_{n}^{\alpha}$  is the fractal energy eigenvalue corresponding to the \(n\)th energy level. The solution of Eq.\eqref{uffupo} the fractal radial wave function can be expressed as:
\begin{align}
R_{n\ell}^{\alpha}(r) &= A_{n\ell} \left(\frac{2}{nS_{\mathbf{F}}^{\alpha}(a_0)}\right)^{\ell+1} \exp\left(\frac{-S_{\mathbf{F}}^{\alpha}(r)}{nS_{F}^{\alpha}(a_0)}\right) \left(L_{n-\ell-1}^{2\ell+1}\left
(\frac{2S_{\mathbf{F}}^{\alpha}(r)}{nS_{\mathbf{F}}
^{\alpha}(a_0)}\right)\right)\nonumber\\
&\propto A_{n\ell} \left(\frac{2}{na_0^{\alpha}}\right)^{\ell+1} \exp\left(\frac{-r^{\alpha}}{na_0^{\alpha}}\right) \left(L_{n-\ell-1}^{2\ell+1}
\left(\frac{2r^{\alpha}}{na_0^{\alpha}}\right)\right),
\end{align}
where $ L_{n-\ell-1}^{2\ell+1}$ represents the associated Laguerre polynomial. The complete fractal wave function is given by the product of the radial and angular parts:
\begin{equation}
  \psi_{n, l, m}^{\alpha}(r, \theta, \phi)=R_{n\ell}^{\alpha}(r)(r)Y_{l, m}(\theta, \phi),
\end{equation}
where $Y_{l, m}(\theta, \phi)$  spherical harmonics that determine the angular behavior. For example, the fractal orbital function for the $1s$ state of the hydrogen atom in spherical coordinates is given by
\begin{align}
  \psi_{10}(r, \theta, \phi) &= R_{10}(r) \cdot Y_{0,0}(\theta, \phi)\\
 &= \frac{2}{\sqrt{4\pi}} \left(\frac{1}{S_{F}^{\alpha}(a_0)^3}\right)^{1/2} \exp\left(\frac{-S_{\mathbf{F}}^{\alpha}(r)}{S_{\mathbf{F}}^{\alpha}(a_0)^3}\right)\\
&\propto \frac{2}{\sqrt{4\pi}}  \left(\frac{1}{a_0^{3\alpha}}\right)^{1/2} \exp\left(\frac{-r^{\alpha}}{a_0^{3\alpha}}\right).
\end{align}
\section*{Probability Density for Radial Hydrogen Atom}
\begin{figure}[H]
    \centering
    \begin{subfigure}{0.45\textwidth}
        \includegraphics[width=\textwidth]{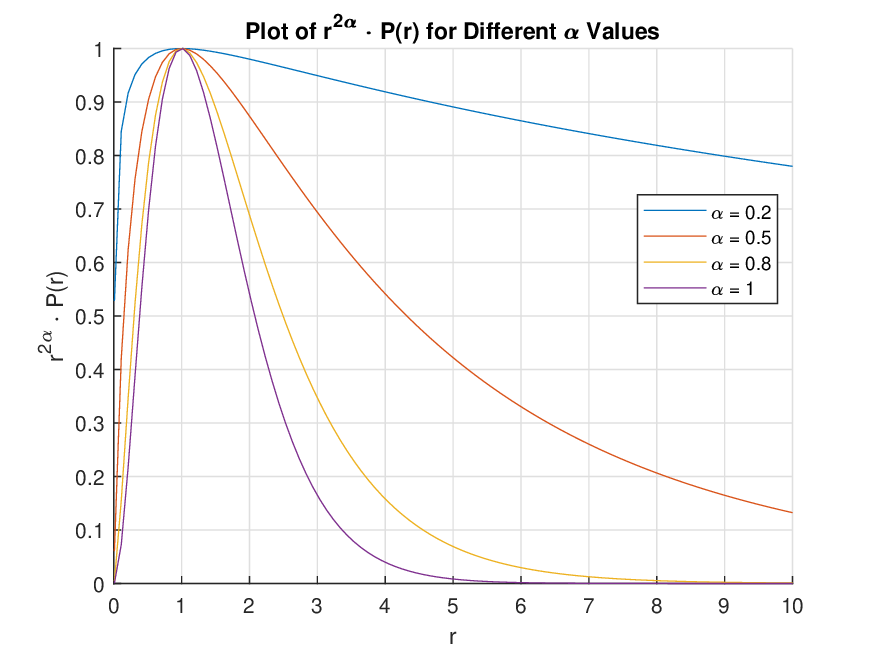}
        \caption{Graph of Eq.\eqref{yyy} for the $n=1,~l=0,~a_{0}=1,~A_{n\ell}=1$.}\label{o221}
   \end{subfigure}
   \hfill
   \begin{subfigure}{0.45\textwidth}
      \includegraphics[width=\textwidth]{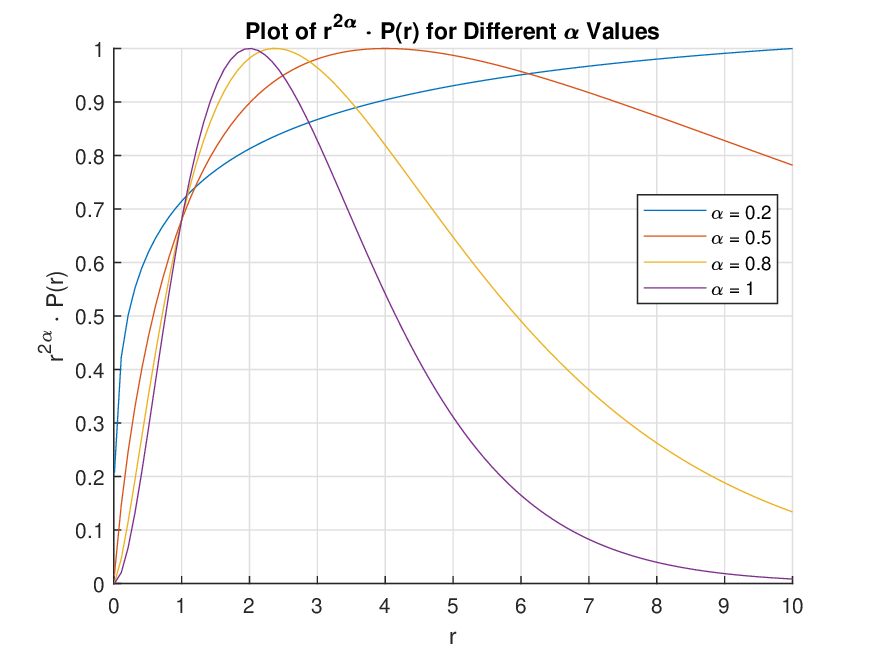}
        \caption{Graph of Eq.\eqref{yyy} for the $n=2,~l=1,~a_{0}=1,~A_{n\ell}=1$.}\label{3o221}
    \end{subfigure}

   \vspace{1em}

    \begin{subfigure}{0.45\textwidth}
       \includegraphics[width=\textwidth]{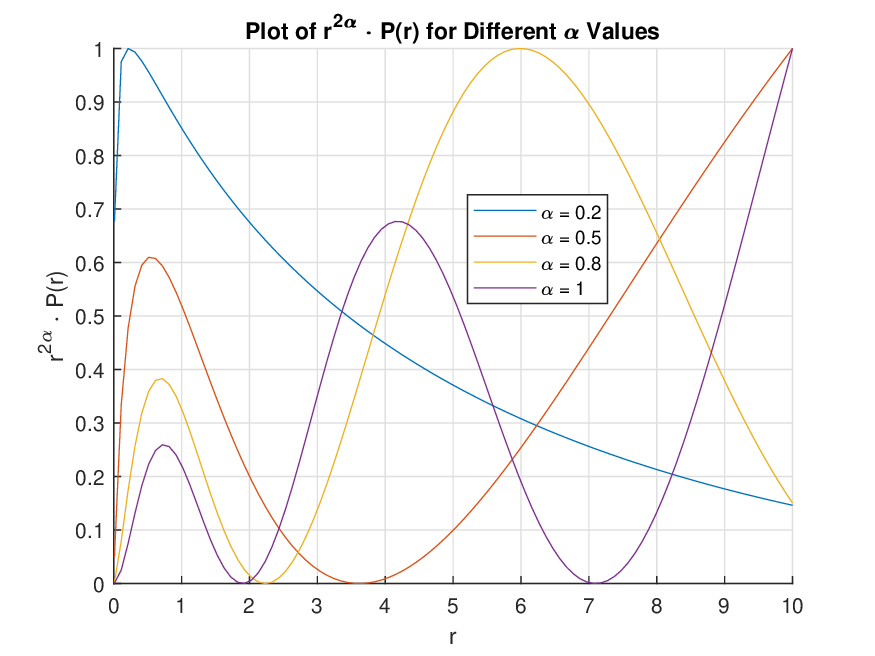}
       \caption{Graph of Eq.\eqref{yyy} for the $n=3,~l=0,~a_{0}=1,~A_{n\ell}=1$.}\label{o4221}
   \end{subfigure}
   \hfill
   \begin{subfigure}{0.45\textwidth}
       \includegraphics[width=\textwidth]{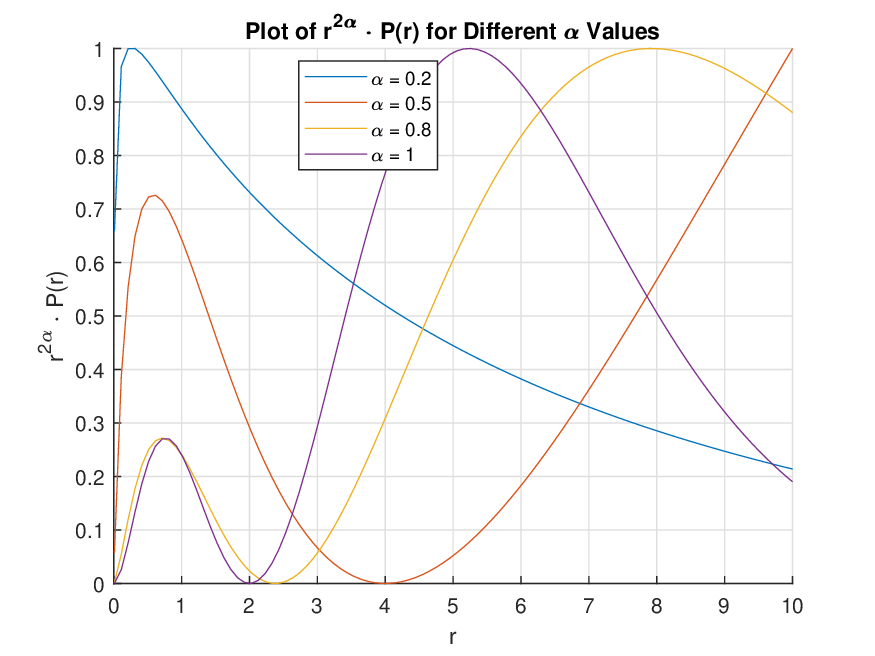}
       \caption{Graph of Eq.\eqref{yyy} for the $n=2,~l=0,~a_{0}=1,~A_{n\ell}=1$.}\label{o42h21}
   \end{subfigure}
   \caption{Probability Density Function of  Hydrogen Atom}
   \label{fig:four_pictures}
\end{figure}

The probability density \(P^{\alpha}(r)\) for the radial part of the hydrogen atom wave function is given by the square of the radial wave function \(R_{n\ell}^{\alpha}(r)\):
\begin{equation}
P^{\alpha}(r) = |\Psi^{\alpha}(\mathbf{r})|^2 = R_{n\ell}^{\alpha}(r)^2.
\end{equation}
In terms of the provided expression for \(R_{n\ell}^{\alpha}(r)\):
\begin{equation}
R_{n\ell}^{\alpha}(r) \propto A_{n\ell} \left(\frac{2}{n a_0^{\alpha}}\right)^{\ell+1} \exp\left(-\frac{ r^{\alpha}}{n a_0^{\alpha}}\right) \left(L_{n-\ell-1}^{2\ell+1}\left(\frac{2 r^{\alpha}}{n a_0^{\alpha}}\right)\right).
\end{equation}
The fractal probability density \(P^{\alpha}(r)\) can be calculated as:
\begin{equation}\label{yyy}
P^{\alpha}(r) = |R_{n\ell}^{\alpha}(r)|^2 = |A_{n\ell}|^2 \left(\frac{2}{n a_0^{\alpha}}\right)^{2(\ell+1)} \exp\left(-\frac{ r^{\alpha}}{n a_0^{\alpha}}\right)  \left|L_{n-\ell-1}^{2\ell+1}\left(\frac{2 r^{\alpha}}{n a_0^{\alpha}}\right)\right|^2.
\end{equation}
This fractal probability density provides information about the distribution of the electron's presence in the hydrogen atom as a function of the radial distance \(r\). In Figures \ref{o221}, \ref{3o221},\ref{o4221}, and \ref{o42h21} we have shown how the fractal dimension changes the fractal probability density function.




\section*{Energy Levels of Hydrogen Atom}
\begin{figure}[H]
  \centering
  \includegraphics[scale=0.5]{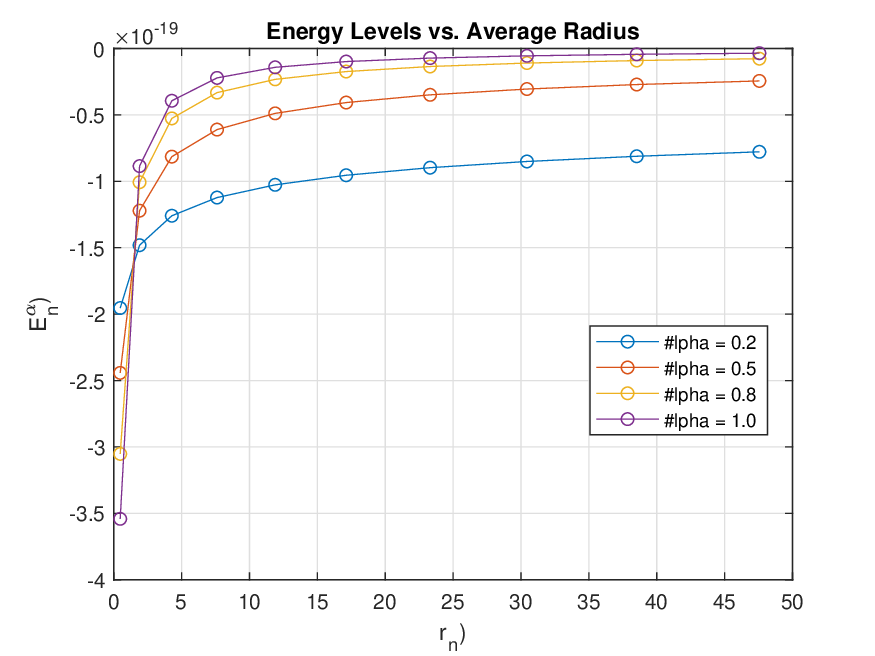}
 \caption{Graph illustrating Eq.\eqref{urr} for different values of the space dimension.}\label{lsso4221}
\end{figure}

Within the Bohr model of the hydrogen atom, energy levels become quantized and are determined by the equation:
\begin{equation}
E_n = -\frac{13.6 \, \text{eV}}{n^2}.
\end{equation}
In this equation, \(n\) represents the principal quantum number. Each energy level \(n\) corresponds to a specific electron orbit or shell, with higher energy levels associated with larger orbits.
The average radius \(r_n\) of the \(n\)th energy level's orbit can be connected to the energy level using the formula:
\begin{equation}
r_n = \frac{n^2  \hbar^2}{  m  e^2}.
\end{equation}
Here, \(n\) denotes the principal quantum number, \(\hbar\) denotes the reduced Planck constant, \(m\) stands for the mass of the electron, and \(e\) represents the elementary charge.
By combining these equations, the energy levels of the hydrogen atom can be expressed in terms of the average radius as:
\begin{equation}\label{ffddde}
E_n = -\frac{m  e^4}{2  \hbar^2}  \frac{1}{r_n}.
\end{equation}
This relationship underscores the inverse relationship between energy levels and the average radius of the electron's orbit. When considering the context of fractal space, we can adapt \eqref{ffddde} as follows:
\begin{align}\label{urr}
  E_n^{\alpha} &= -\frac{ m  e^4}{2 \hbar^2}  \frac{1}{S_{\mathbf{F}}^{\alpha}(r_n)}\nonumber\\
&\propto -\frac{ m  e^4}{2 \hbar^2}  \frac{1}{r_n^\alpha}.
\end{align}
In Figure \ref{lsso4221}, we have visualized the behavior described by Eq.\eqref{urr} to illustrate the impact of the fractal dimension on energy levels within the Hydrogen atom. The figure depicts that as the fractal dimension decreases, the energy levels in the fractal space undergo an upward shift.

\section{Fractal Schr\"{o}dinger Equation of Simple Harmonic Motion }
The fractal time-dependent harmonic Schr\"{o}dinger equation, which includes the time evolution of the wave function, is suggested by:
\begin{equation}\label{fffddaqw}
i\hbar D_{t}^{\beta}\Psi(x, t) = \left(-\frac{\hbar^2}{2m} D_{x}^{2\alpha} + \frac{1}{2} m (\omega^{\alpha})^2 S_{\mathbf{F}}^{\alpha}(x)^2\right) \Psi(x, t),~~~x,t\in \mathbf{F},
\end{equation}
where $\omega^{\alpha}$ is the angular frequency of the fractal harmonic oscillator.
The fractal one-dimensional  time-independent harmonic Schr\"{o}dinger equation is suggested as:
\begin{equation}
-\frac{\hbar^2}{2m} D_{x}^{2\alpha}\psi_{n}^{\alpha}(x) + \frac{1}{2} m (\omega^{\alpha})^2 S_{\mathbf{F}}^{\alpha}(x)^2 \psi_{n}^{\alpha}(x) = E_{n}^{\alpha} \psi_{n}^{\alpha}(x),
\end{equation}
its solution is
\begin{align}
\psi_{n}^{\alpha}(x)&=\frac{1}{\sqrt{2^n n!}} \left(\frac{m \omega^{\alpha}}{\pi \hbar}\right)^{1/4} \exp\left(-\frac{m\omega^{\alpha} S_{\mathbf{F}}^{\alpha}(x)^2}{2\hbar}\right) H_n\left(\sqrt{\frac{m \omega^{\alpha}}{\hbar}} S_{\mathbf{F}}^{\alpha}(x)\right)\nonumber\\&\propto
\frac{1}{\sqrt{2^n n!}} \left(\frac{m \omega^{\alpha}}{\pi \hbar}\right)^{1/4} \exp\left(-\frac{m\omega^{\alpha} x^{2\alpha}}{2\hbar}\right) H_n\left(\sqrt{\frac{m \omega^{\alpha}}{\hbar}} x^{\alpha}\right),
\end{align}
where $H_n$ stands for the $n$-th Hermite polynomial.
Then the time-dependent solution of Eq.\eqref{fffddaqw} is
 \begin{equation}
\Psi(x, t) = \sum_n c_n \psi_{n}^{\alpha}(x) \exp\left(\frac{-iE_n^{\alpha} S_{\mathbf{F}}^{\beta}(t)}{\hbar}\right).
\end{equation}
\section*{Fractal Probability Density Function of  Harmonic Oscillator }
The probability density function for a quantum harmonic oscillator is given by the square of the absolute value of its wavefunction:
\begin{align}\label{p}
  P^{\alpha}(x) = |\psi_n^{\alpha}(x)|^2&= \left| \frac{1}{\sqrt{2^n n!}} \left(\frac{m \omega^{\alpha}}{\pi \hbar}\right)^{1/4} \exp\left(-\frac{m\omega^{\alpha} S_{\mathbf{F}}^{\alpha}(x)^2}{2\hbar}\right) H_n\left(\sqrt{\frac{m \omega^{\alpha}}{\hbar}} S_{\mathbf{F}}^{\alpha}(x)\right) \right|^2\nonumber\\&
\propto \left| \frac{1}{\sqrt{2^n n!}} \left(\frac{m \omega^{\alpha}}{\pi \hbar}\right)^{1/4} \exp\left(-\frac{m\omega^{\alpha} x^{2\alpha}}{2\hbar}\right) H_n\left(\sqrt{\frac{m \omega^{\alpha}}{\hbar}} x^{\alpha}\right) \right|^2.
\end{align}
In Figure \ref{fig:four_pictures}, we've depicted the probability density function of simple harmonic motion. This visualization effectively illustrates how the wavefunction is influenced by the dimensionality of fractal space.
\begin{figure}[H]
    \centering
    \begin{subfigure}{0.45\textwidth}
        \includegraphics[width=\textwidth]{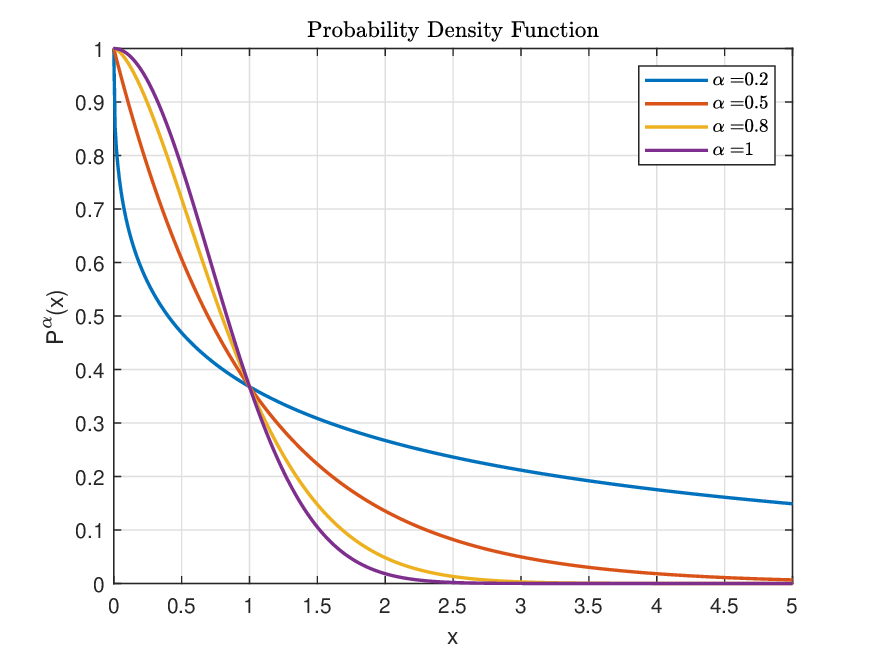}
        \caption{Caption for Image n=0}
    \end{subfigure}
    \hfill
    \begin{subfigure}{0.45\textwidth}
        \includegraphics[width=\textwidth]{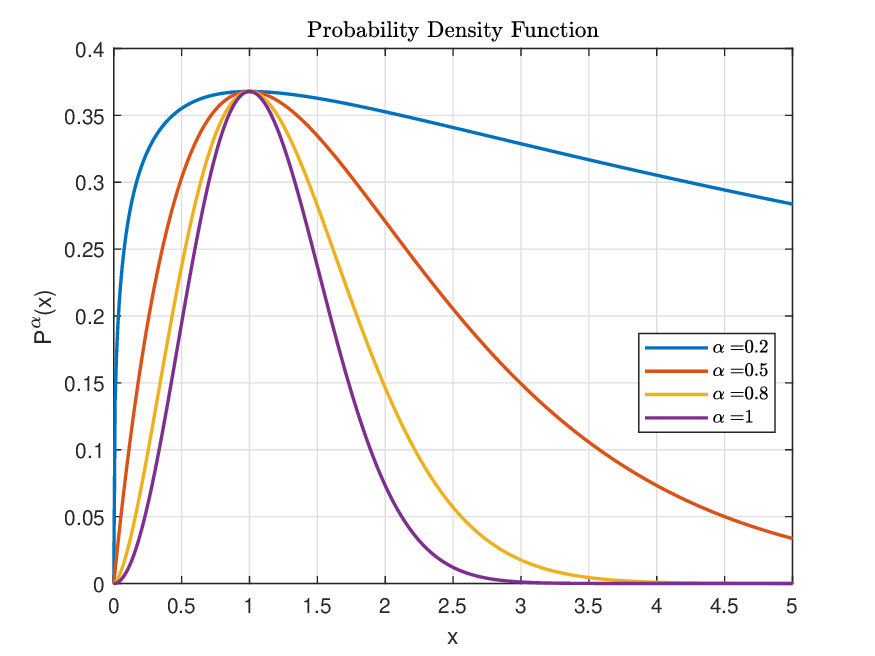}
        \caption{Caption for Image n=1}
    \end{subfigure}

    \vspace{1em}

    \begin{subfigure}{0.45\textwidth}
        \includegraphics[width=\textwidth]{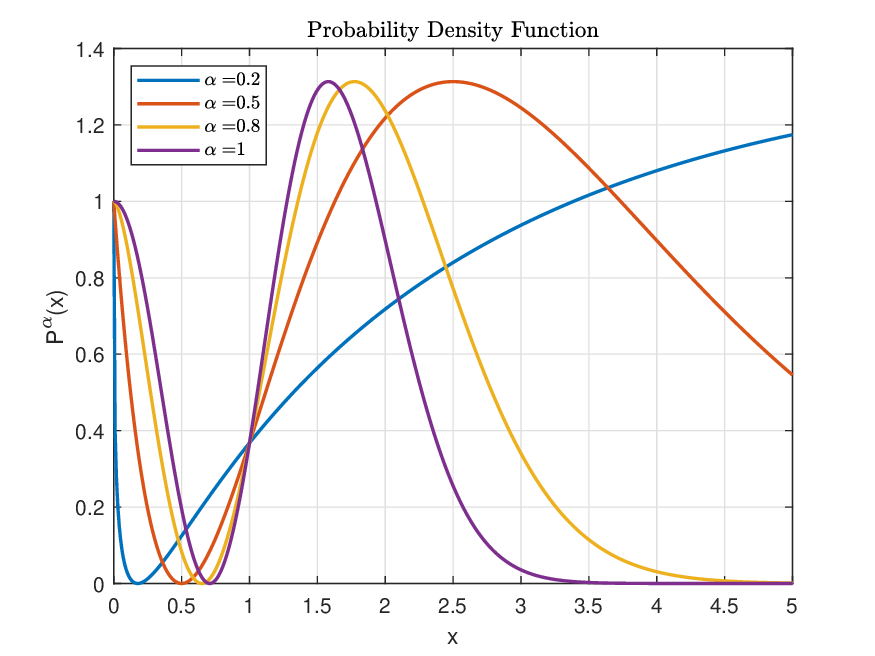}
        \caption{Caption for Image n=2}
    \end{subfigure}
    \hfill
    \begin{subfigure}{0.45\textwidth}
        \includegraphics[width=\textwidth]{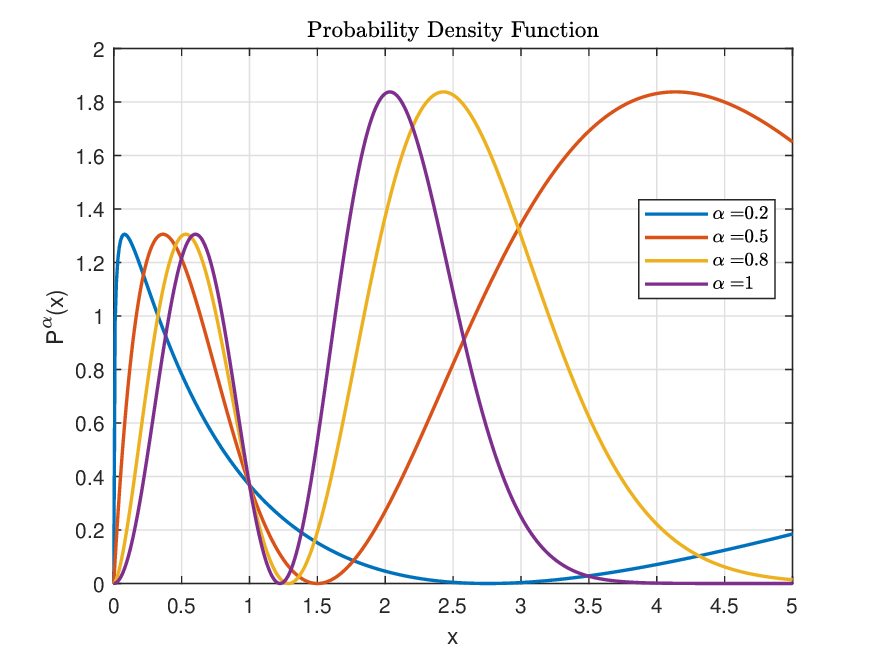}
        \caption{Caption for Image n=3}
    \end{subfigure}
    \caption{Probability Density Function}
    \label{fig:four_pictures}
\end{figure}





\section*{Fractal Energy Levels of Simple Harmonic Oscillator}
In the realm of quantum mechanics, the wave-like nature of particles gives rise to a fascinating phenomenon: energy levels become quantized. This signifies that the energy associated with a quantum harmonic oscillator can exclusively adopt specific, distinct values. These discrete energy levels unveil a remarkable feature when viewed through the lens of fractal space, as outlined below:\\
The energy levels in this context are denoted as $E_n^{\alpha} $ and are coined "fractal energy levels." They take the form:
\begin{align}\label{kkiio}
E_n^{\alpha} &= \hbar \omega^{\alpha} \left(n + \frac{1}{2}\right)
,~~~n=0,1,2,...
\end{align}
Another representation arises from the exploration of fractal space, yielding:
\begin{align}
  E_n^{\alpha}&=\hbar \sqrt{\frac{S_{F}^{\alpha}(x)}{m}} \left(n + \frac{1}{2}\right)\nonumber\\&
\propto \hbar \sqrt{\frac{x^{\alpha}}{m}} \left(n + \frac{1}{2}\right).
\end{align}
This intriguing relationship is linked to the fractal harmonic oscillator's angular frequency $\omega^{\alpha} $.

\begin{figure}[H]
  \centering
  \includegraphics[scale=0.5]{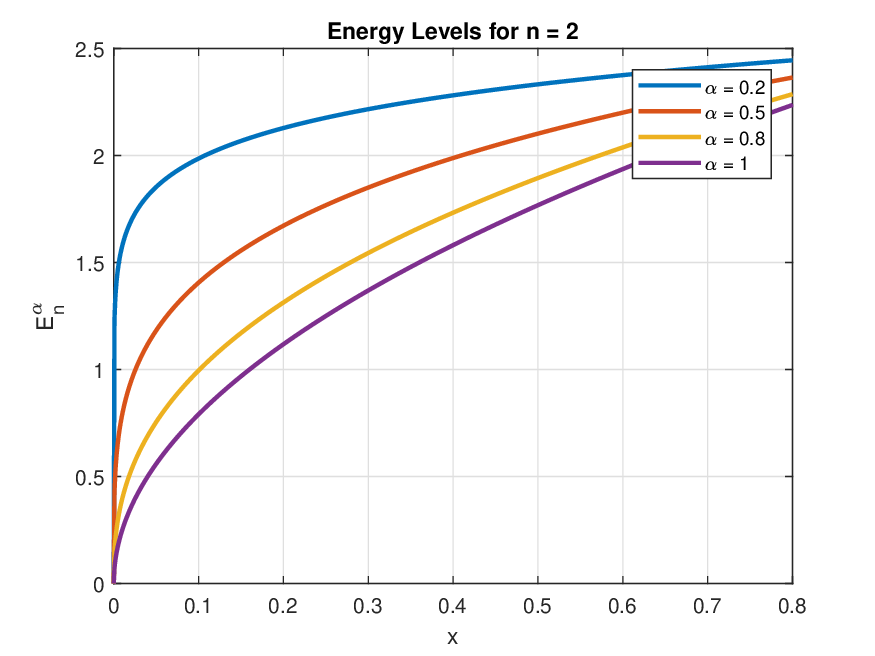}
  \caption{Graph of Eq.\eqref{kkiio}}\label{d2m21}
\end{figure}
 Figure \ref{d2m21} reveals that as the fractal dimension of space decreases, the energy level within the harmonic oscillator increases.
\section{Conclusion \label{3g}}
In conclusion, this study demonstrates the profound interconnection between fractal calculus and quantum mechanics. By embarking on an exploration of fractal sets, we have illuminated the applications and implications of the Fractal Schr\"{o}dinger Equation. The unveiled General Solution for the Time-Dependent Schr\"{o}dinger Equation provides a foundational understanding of its dynamics in the fractal realm. Analyzing the Probability Density of the Radial Hydrogen Atom within the fractal framework reveals the intricate influence of dimensionality on quantum behavior. Furthermore, the investigation into the Energy Levels of the Hydrogen Atom establishes a bridge between fractal geometry and quantum states. The extension of the Fractal Schr\"{o}dinger Equation to Simple Harmonic Motion introduces the Fractal Probability Density Function for the Harmonic Oscillator, expanding our insights into the quantum realm within a fractal context. The visual aids incorporated throughout the research enhance the clarity of the findings. In synthesizing quantum mechanics and fractal mathematics, this study not only unravels their synergy but also lays the groundwork for deeper explorations at the nexus of these captivating fields.\\
\textbf{Declaration of Competing Interest:}\\
The authors declare that they have no known competing financial interests or personal relationships that could have appeared to influence the work reported in this paper.\\
\textbf{Declaration of generative AI and AI-assisted technologies in the writing process.}
During the preparation of this work, the authors used GPT in order to correct grammar and writing. After using this GPT, the authors reviewed and edited the content as needed and takes full responsibility for the content of the publication.

 \bibliographystyle{elsarticle-num}
 \bibliography{Refrancesma3}





\end{document}